Superconductivity in CeOBiS$_2$ with cerium valence fluctuation

Masanori Nagao[a,*], Akira Miura[b], Ikuo Ueta[a], Satoshi Watauchi[a], and Isao Tanaka[a]

[a]*University of Yamanashi, 7-32 Miyamae, Kofu, Yamanashi 400-8511, Japan*

[b]*Hokkaido University, Kita13 Nishi8, Kita-ku, Sapporo, Hokkaido 060-8628, Japan*

[*]Corresponding Author

Masanori Nagao

Postal address: University of Yamanashi, Center for Crystal Science and Technology

Miyamae 7-32, Kofu 400-8511, Japan

Telephone number: (+81)55-220-8610

Fax number: (+81)55-254-3035

E-mail address: mnagao@yamanashi.ac.jp

**Abstract**

Resistivities of single-crystalline as well as poly-crystalline samples of $CeOBiS_2$ without fluorine doping were measured at temperatures down to 0.13 K, and were compared with those of poly-crystalline $LaOBiS_2$ and $PrOBiS_2$. Both poly-crystalline and single-crystalline $CeOBiS_2$ exhibited zero resistivity below 1.2 K while poly-crystalline $LaOBiS_2$ and $PrOBiS_2$ did not show zero resistivity down to 0.13 K. Superconducting transition temperature of $CeOBiS_2$ was reduced by increasing the applied current density. The superconductivity of $CeOBiS_2$ without chemical doping is likely triggered by the carriers induced by the valence fluctuation between $Ce^{3+}$ and $Ce^{4+}$.

**Keywords**

A. $BiS_2$-based superconductor; D. Superconductivity; Valence fluctuation; E. Adiabatic demagnetization refrigerator;

**Main text**

**1. Introduction**

  Chemical doping of semiconductors is a method used to induce superconductivity in them. The superconductivity of diamond [1,2], Si [3], and SiC [4-6] is induced by introducing carriers through chemical doping. $R$(O,F)BiS$_2$ ($R$ = La, Ce, Pr, Nd, Yb) [7-11] are layered semiconductors with a band gap of around 1 eV [12], and superconductivity is induced in them through electron carrier doping by substituting oxygen (O) with fluorine (F). This doping induces electron carriers into the BiS$_2$ superconducting layer. Recently, it was discovered that the fluctuation of Eu valence between Eu$^{2+}$ and Eu$^{3+}$ induces carriers and produces superconductivity in EuFBiS$_2$ and Eu$_3$Bi$_2$S$_4$F$_4$ without chemical doping [13,14]. Thus, Eu-based compounds are examples in which superconductivity is induced by the fluctuation of valence. As suggested in Ref. 15 and 16, for $R$ = Ce, the Ce valence in CeOBiS$_2$ also fluctuates between Ce$^{3+}$ and Ce$^{4+}$. However, its zero resistivity has not been reported at a temperature down to 1.8 K in F-free CeOBiS$_2$. This motivated us to examine superconductivity in non-chemically doped CeOBiS$_2$ to investigate whether a valence fluctuation can induce superconductivity in general. In this paper, we measured the temperature dependence of electrical resistivity for chemically non-doped $R$OBiS$_2$ ($R$ = La, Ce, Pr) poly crystals and CeOBiS$_2$ single crystals down to 0.13 K using an adiabatic demagnetization refrigerator (ADR). In contrast to LaOBiS$_2$ and PrOBiS$_2$, only CeOBiS$_2$ showed superconducting transition below 1.2 K, indicating that the valence fluctuation of Ce induces superconductivity.

## 2. Experimental

Poly-crystalline samples of $R$OBiS$_2$ ($R$ = La, Ce, Pr) were synthesized by a solid state reaction in a vacuumed quartz tube using $R_2$S$_3$ (99.9 wt%), Bi$_2$S$_3$ (99.9 wt%), and Bi$_2$O$_3$ (99.9 wt%) as raw materials. The raw materials were mixed in a nominal composition of $R$OBiS$_2$ using a mortar, and sealed into a quartz tube in vacuum (~10 Pa). Then, the sample was heated at 700 °C for 10 h. The calcined mixture was ground to homogenize it and pressed into a pellet of 10 mm diameter and about 1-2 mm thickness, and then sealed in a quartz tube under vacuum (~10 Pa). This pellet in the quartz tube was heated at 700 °C for 20 h, and then furnace-cooled to room temperature. The quartz tube was opened in air. Poly-crystalline samples of $R$OBiS$_2$ were obtained as products.

Single crystals of CeOBiS$_2$ were grown by the CsCl flux method in a vacuumed quartz tube using $Ce_2S_3$ (99.9 wt%), $Bi_2S_3$ (99.9 wt%), $Bi_2O_3$ (99.9 wt%), and CsCl (99.8 wt%) as raw materials [17-20]. The raw materials were weighed to obtain the nominal composition of $CeOBiS_2$. A mixture of these raw materials (0.8 g) and CsCl flux (5.0 g) was combined using a mortar, and then sealed in a quartz tube under vacuum (~10 Pa). This mixed powder in the quartz tube was heated at 950 °C for 10 h, cooled slowly to 650 °C at a rate of 1 °C/h, and then furnace-cooled to room temperature. The quartz tube was opened in air, and the flux was dissolved in the quartz tube using distilled water. The product obtained was filtered and washed with distilled water.

The crystal structures of the poly-crystalline and single-crystalline products were evaluated by X-ray diffraction (XRD) analysis using CuK$\alpha$ radiation. The resistivity–temperature ($\rho$–$T$) characteristics of these poly-crystalline and single-crystalline samples were measured by the standard four-probe method at a constant current density ($J$) using an ADR option for quantum design physical property measurement system (PPMS). The magnetic field applied for operating the ADR was 3 T at 1.9 K; subsequently, it was removed. Consequently, the temperature of sample decreased to around 0.13 K. The measurement of $\rho$–$T$ characteristics was started at the lowest temperature (around 0.13 K), which was spontaneously increased to around 15 K. The $\rho$–$T$ characteristics were measured while the temperature was increased. The dependence of applied current density on the $\rho$–$T$ characteristics was also observed. The superconducting transition temperature ($T_c$) was estimated from the $\rho$–$T$ characteristics. The transition temperature corresponding to the onset of superconductivity ($T_c^{onset}$) is defined as the temperature at which deviation from linear behavior is observed in the normal conducting state in the $\rho$–$T$ characteristics. The zero resistivity ($T_c^{zero}$) is determined as the temperature at which resistivity is below 5 $\mu\Omega$ cm. $\rho$–$T$ characteristics under a magnetic field ($H$) range of 0-0.3 T parallel to the $ab$-plane and the $c$-axis of CeOBiS$_2$ single crystal were measured without ADR option in the temperature range of 2.0-5.0 K.

## 3. Results and discussion

Figure 1 shows the XRD patterns of $R$OBiS$_2$ poly-crystalline samples where (a) $R$ = La, (b) $R$ = Ce, and (c) $R$ = Pr. The major diffraction peaks of the obtained poly-crystalline samples were indexed as $R$OBiS$_2$ phases. Figure 2 shows the $\rho$–$T$ characteristics of $R$OBiS$_2$ poly-crystalline samples in the temperature range of 0.13-15 K. LaOBiS$_2$ and PrOBiS$_2$ poly-crystalline samples did not exhibit zero resistivity down to around 0.13 K. In contrast, for $R$ = Ce, the resistivity drops in the range 1.3-1.9 K; the $T_c^{onset}$ and $T_c^{zero}$ were estimated to be 1.9 K and 1.3 K, respectively. The transport properties of the $R$OBiS$_2$ poly-crystalline samples exhibited semiconducting behavior. LaOBiS$_2$ showed a significantly higher resistivity compared to CeOBiS$_2$ and PrOBiS$_2$. The resistivity of PrOBiS$_2$ was lower than that of CeOBiS$_2$ at the normal state. However, superconducting transition was not observed in PrOBiS$_2$ down to 0.13 K.

Figure 3 shows the XRD pattern of a well-developed plane of CeOBiS$_2$ single crystal. The presence of only 00$l$ diffraction peaks of the CeOBiS$_2$ structure indicates that the $ab$-plane is well developed. Four terminals for the $\rho$–$T$ characteristics measurement were fabricated on the well-developed plane using silver paste. In consequence, the applied current was along the $ab$-plane. The $\rho$–$T$ characteristics of the CeOBiS$_2$ single crystal sample were measured in the temperature range of 0.13-15 K. Figure 4 shows the applied current density dependence of $\rho$–$T$ characteristics for a CeOBiS$_2$ single crystal. Superconducting transition was observed in the CeOBiS$_2$ single crystal, and the superconducting transition temperature was reduced by increasing the applied current density ($J$). The $T_c^{zero}$ was not observed down to 0.13 K, when $J$ was more than 6.33 A/cm$^2$. For $J$ = 1.05 A/cm$^2$, the $T_c^{onset}$ and $T_c^{zero}$ were estimated to be 3.1 K and 1.2 K, respectively. Similar $T_c^{zero}$ found in poly-crystalline and single-crystalline samples suggests that CeOBiS$_2$ without intentional doping displays superconductivity below 1.2 K. Nonetheless, there are minor differences in transition temperatures between poly-crystalline and single-crystalline samples. The $T_c^{zero}$ of single-crystal sample (1.2 K) was slightly lower than that of the poly-crystalline sample (1.3 K) because of the high applied current density associated with the smaller cross-section area of a single crystal. The $T_c^{onset}$ and $\Delta T_c$ ($T_c^{onset}$ - $T_c^{zero}$) of single-crystalline sample increased in comparison with the poly-crystalline sample. We cannot deny the possibility of contamination by F detected in the CsCl flux, because the fluorine content in CsCl was determined to be 0.11 mol% by ion chromatography analysis. Nonetheless, F detection in CeOBiS$_2$ single crystals was very difficult by spectroscopic analysis. For spectroscopic analysis such as electron probe microanalysis (EPMA), F-K$_\alpha$ (677 eV) characteristic X-ray is overlapped with Ce-M$_\zeta$ (676 eV) characteristic X-ray [21]. On the other hand, for ion chromatography analysis, dissolution of the CeOBiS$_2$ single crystals into a solution was found to be difficult.

The $\rho$–$T$ characteristics for the CeOBiS$_2$ single crystal in the range of 2.0-5.0 K under a magnetic field ($H$) range of 0-0.3 T parallel to the *ab*-plane and *c*-axis using PPMS without ADR option are shown in Figure 5. The resistivity dropped at around 3.5 K by the self-field ($H$ = 0 T), and the temperature at which resistivity dropped decreased with the increase in magnetic field ($H$). This result is evidence that superconducting transition ($T_c^{onset}$) occurs at around 3.5 K. This $T_c^{onset}$ is higher than that measured using

ADR option (3.1 K). This phenomenon may be explained as follows: the measurement of $T_c^{onset}$ without ADR option was performed in zero field cooling; on the other hand, the measurement with ADR option was performed with field cooling. The decrease in $T_c^{onset}$ under the magnetic field applied parallel to the $c$-axis is more significant than that of the $ab$-plane, which corresponds to F-doped $CeOBiS_2$ single crystals [20].

In both poly-crystalline and single-crystalline samples, $CeOBiS_2$ shows zero resistivity below 1.2 K. We assumed that the carrier for initiating superconductivity in $CeOBiS_2$ originated from the valence fluctuations between $Ce^{3+}$ and $Ce^{4+}$ states [22]. This is further supported by the absence of superconducting transition down to 0.13 K in $LaOBiS_2$ and $PrOBiS_2$. The valence of La is probably $La^{3+}$ in $LaOBiS_2$. In the case of $PrOBiS_2$, it was observed using X-ray photoelectron spectroscopy (XPS) analysis and magnetization-temperature ($M$-$T$) characterization that valence fluctuations between $Pr^{3+}$ and $Pr^{4+}$ states hardly exist [23]. Therefore, the multivalence state of $Ce^{3+}$ and $Ce^{4+}$ induced superconductivity in $CeOBiS_2$ without chemically doping it.

## 4. Conclusion

The superconductivity of $CeOBiS_2$ without fluorine doping was examined down to 0.13 K. Poly-crystalline $CeOBiS_2$ showed zero resistivity below 1.3 K while $LaOBiS_2$ and PrOBiS$_2$ poly-crystalline samples did not show superconductivity down to 0.13 K. The superconductivity of the CeOBiS$_2$ single crystal was also observed at around 1.2 K, and it was similar to that found in the poly-crystalline sample. Therefore, the valence fluctuations between Ce$^{3+}$ and Ce$^{4+}$ states can induce superconducting transition without chemical doping, showing the potential of such valence fluctuations for exploring new superconductors.

**Acknowledgments**

This work was supported by JSPS KAKENHI (Grant-in-Aid for challenging Exploratory Research) Grant Number 15K14113.

We would like to thank Editage (www.editage.jp) for English language editing.

**Figure captions**

Figure 1. XRD pattern of $R$OBiS$_2$ poly-crystalline samples where (a) $R$ = La, (b) $R$ = Ce and (c) $R$ = Pr.

Figure 2. $\rho$–$T$ characteristics of $R$OBiS$_2$ poly-crystalline samples at 0.13-15 K using ADR option.

Figure 3. XRD pattern of well-developed plane of CeOBiS$_2$ single crystal.

Figure 4. Applied current density dependence of $\rho$–$T$ characteristics for the CeOBiS$_2$ single crystal at 0.13-15 K using ADR option.

Figure 5. $\rho$–$T$ characteristics for the CeOBiS$_2$ single crystal at 2.0-5.0 K under a magnetic field ($H$) of 0-0.3 T parallel to the (a) *ab*-plane and (b) *c*-axis without ADR option.

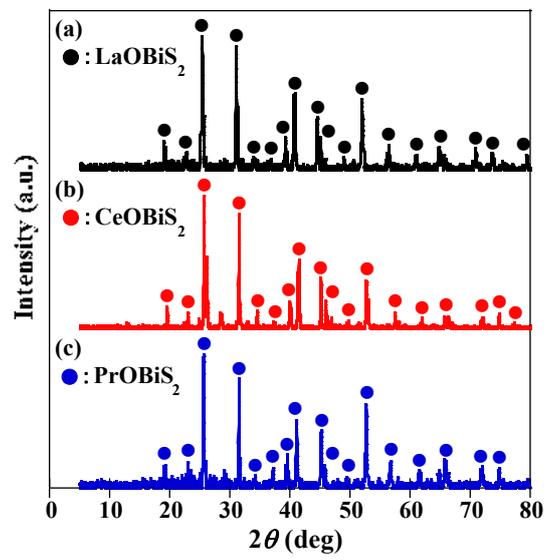

**Figure 1**

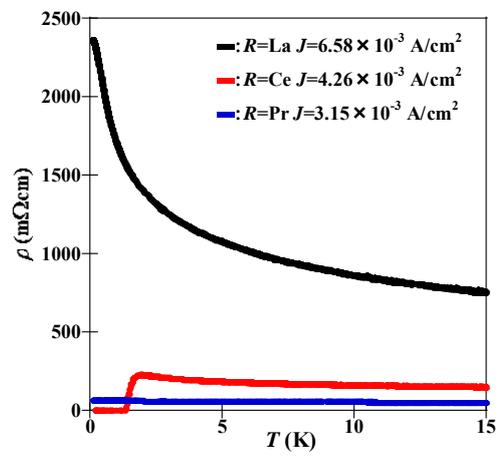

**Figure 2**

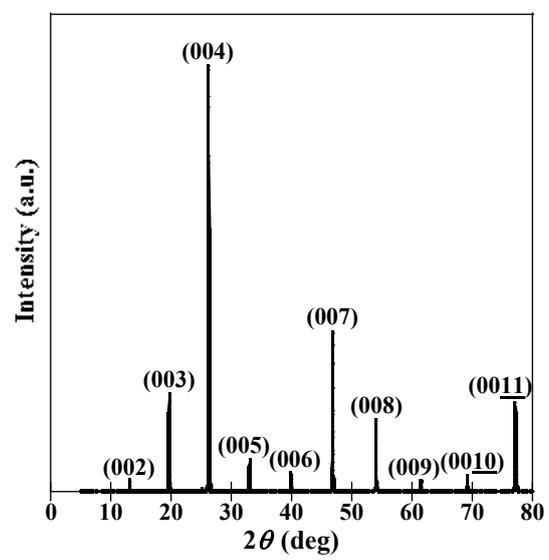

**Figure 3**

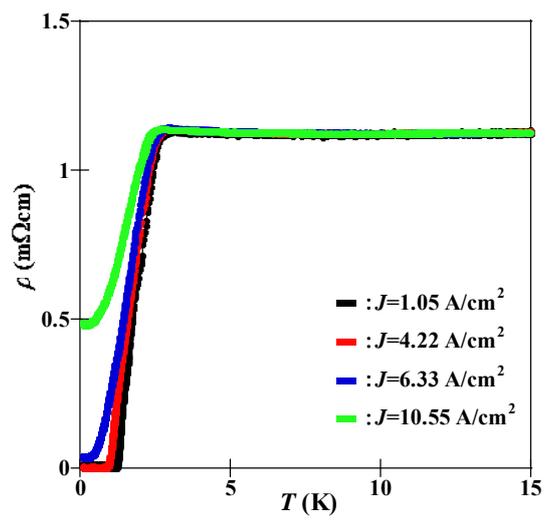

**Figure 4**

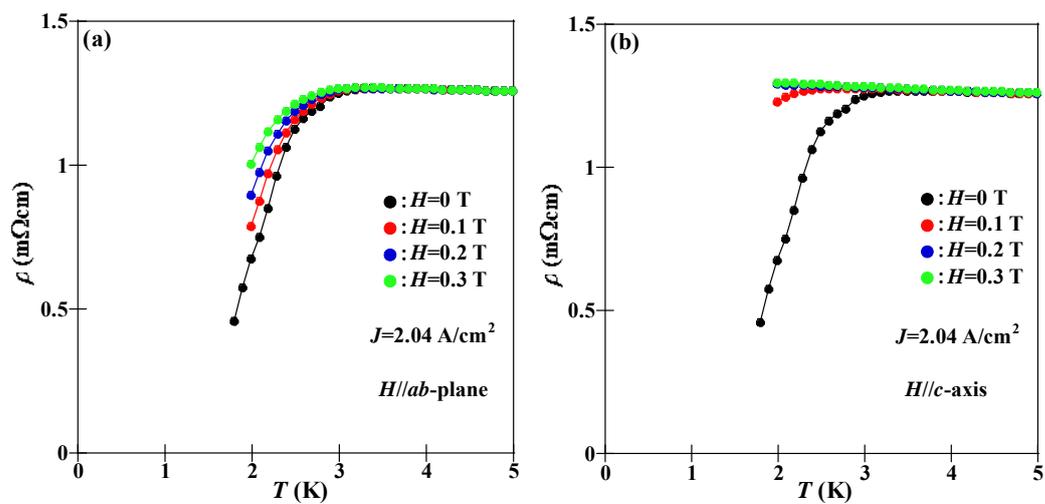

**Figure 5**